\documentclass[pre,twocolumn,aps,showpacs]{revtex4}
\newcommand{\bec}[1]{\mbox{\boldmath $ #1$}}
\usepackage{graphicx}
\begin{document}

\title{Mixing at the external boundary of a submerged turbulent jet}
\medskip
\author{A. Eidelman$^{a}$}
\email{eidel@bgu.ac.il}
\author{T. Elperin$^{a}$}
\email{elperin@bgu.ac.il}
\homepage{http://www.bgu.ac.il/~elperin}
\author{N. Kleeorin$^a$}
\email{nat@bgu.ac.il}
\author{G.~Hazak$^b$}
\email{giohazak@netvision.net.il}
\author{I. Rogachevskii$^a$}
\email{gary@bgu.ac.il}
\homepage{http://www.bgu.ac.il/~gary}
\author{O.~Sadot$^b$}
\email{sorens@bgu.ac.il}
\author{I.~Sapir-Katiraie$^a$}
\email{katiraie@bgu.ac.il}
  \medskip
\affiliation{$^a$The Pearlstone Center for Aeronautical Engineering
 Studies, Department of Mechanical Engineering,
 Ben-Gurion University of the Negev,
 P. O. Box 653, Beer-Sheva
84105, Israel\\
$^b$Department of Physics, Nuclear Research Center, Negev POB 9001,
Beer-Sheva 84190, Israel}
%\date{\today}
\date{3 October 2008}
\begin{abstract}
We study experimentally and theoretically mixing at the external boundary of a submerged turbulent jet. In the experimental study we use Particle Image Velocimetry and an Image Processing Technique based on the analysis of the intensity of the Mie scattering to determine the spatial distribution of tracer
particles. An air jet is seeded with the incense smoke particles
which are characterized by large Schmidt number and small Stokes
number. We determine the spatial distributions of
the jet fluid characterized by a high concentration of the particles
and of the ambient fluid characterized by a low concentration of
the tracer particles. In the data analysis we use two approaches, whereby one approach is based on the measured phase function for the study of the mixed state of two fluids. The other approach is based on the analysis of the two-point second-order correlation function of the particle number density fluctuations generated by tangling of the gradient of the mean particle number density by the turbulent velocity field. This gradient is formed at the external boundary of a submerged turbulent jet. We demonstrate that PDF of the phase function of a jet fluid penetrating into an external flow and the two-point second-order correlation function of the particle number density do not have universal scaling and cannot be described by a power-law function. The theoretical predictions made in this study are in a qualitative agreement with the obtained experimental results.
\end{abstract}

\pacs{47.27.wg, 47.27.tb}

\maketitle

\section{Introduction}

Jets are often used in numerical, experimental and theoretical studies for
modelling of mixing in various engineering applications (see, e.g.,
\cite{CR73,MB88,HH89,Q89,MB90,HH93,MB95,GG99,GA02,RP03,DIM05}).
Different kinds of structures are formed during mixing in turbulent jets (see, e.g., \cite{CC71,DIM83}). One of the most significant results is a
detection of a sharp increase of mixing rates caused by the onset of
small-scale turbulence within large-scale coherent motions \cite{HZ80}.
This effect that was studied experimentally and numerically,
demonstrates complex nonlinear dynamics of mixing (see, e.g., \cite{H90,MR91,GG99,DIM00,DIM05,MDL06}).  However, it is not
clear how such mixing states are attained, and the ambiguity is
exacerbated by the differences in experimental results.

The possible reason for such variety of experimental results is the difference in Schmidt number between gases and liquids. It was found in \cite{DIM00} that mixing becomes more effective in a liquid jet for the larger Reynolds numbers. However, the dependence of the normalized scalar variance on Reynolds number for the gaseous jet is very weak. It was suggested in \cite{DIM00} that this difference is caused by the Schmidt number effect. In particular, when Schmidt number Sc $\sim 1$ the larger molecular species diffusivity results in better mixing of the scalar field even at lower Reynolds numbers. Here Sc$= \nu / D_m$ is the Schmidt number, $D_m$ is the coefficient of Brownian (molecular) diffusion and $\nu$ is the kinematic viscosity. For the liquid jets, Sc $\sim 10^2$, the improved mixing requires the enhancement in the interfacial surface-to-volume ratio (smaller distances between iso-scalar surfaces) associated with higher Reynolds numbers. Due to the large differences in Schmidt numbers the results of these studies in liquids are not directly applicable to the case of gaseous shear-layer mixing.

In the present study we investigate experimentally mixing at the air jet interface using the incense smoke that is characterized by significantly larger Schmidt number (Sc$\sim 10^5)$ than that employed in the previous studies of
gas flow mixing. The high magnitude of Schmidt number prevents from fast molecular diffusion to affect mixing. We study mixing at the external boundary of a submerged turbulent jet using Particle Image Velocimetry (PIV) and an Image Processing Technique based on the analysis of the intensity of the Mie scattering in order to determine the spatial distribution of the jet fluid (with a high concentration of the tracer particles) and of the ambient fluid (with a low concentration of the tracer particles). In this flow there is no a sharp interface between the jet and ambient air.

In the data analysis we use two approaches, whereby one approach is similar to that used previously in the analysis of Rayleigh-Taylor instability (see \cite{H06}). This approach is based on the measured phase function for the study of the mixed state of two fluids.
The other approach used in our study is based on the analysis of the two-point second-order correlation function of the particle number density fluctuations generated by tangling of the gradient of the mean particle number density by the turbulent velocity field (see, e.g., \cite{EKR95,EKR02}). This gradient is formed at the external boundary of a submerged turbulent jet. Both approaches demonstrate that probability density function (PDF) of the phase function of a jet fluid penetrating into an external flow and the two-point second-order correlation function of the particle number density do not have universal scaling and cannot be fitted by a power-law function. There is a qualitative agreement between theoretical and experimental results obtained in our study.

The paper is organized as follows. In Sect.~II we describe the experimental set-up for a laboratory study of mixing at the external boundary of a submerged turbulent jet. In Sect.~III we discuss the experimental results and their detailed analysis by means of the approach based on the measured phase function. In Sect.~IV we perform a theoretical study of mixing at the external boundary of a submerged turbulent jet based on the analysis of the two-point second-order correlation function of the particle number density. In Sect.~IV we also compare the theoretical predictions with the obtained experimental results for the correlation function of the particle number density. Finally, conclusions are drawn in Sect.~V. In Appendix we perform an asymptotic analysis of the solution of equation for the two-point second-order correlation function.

\section{Experimental set-up}

\begin{figure}
\vspace*{2mm} \centering
\includegraphics[width=8cm]{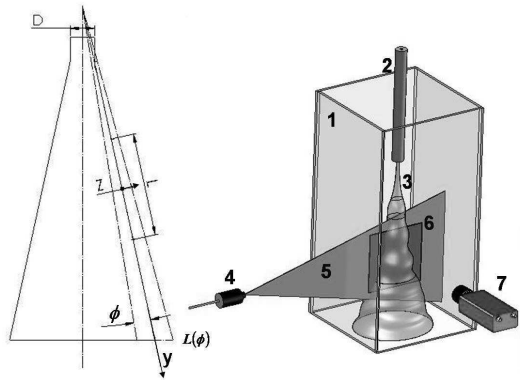}
\caption{\label{Fig1} A scheme of the test section and PIV technique; a sketch of a jet coordinates and a range of measurements.
1 - channel with transparent walls, 2 - tube with a jet nozzle, 3 - submerged jet,
4 - light sheet optics, 5 - laser light sheet, 6 - image area, 7 - CCD camera.}
\end{figure}

The experimental set-up with air jet shown in Fig.~\ref{Fig1} includes the chamber
with the transparent plexiglass walls (1), and the cylindrical tube
(2) with a jet nozzle mounted at the tip. A
submerged air jet (3) discharges into the chamber. The cross-section of the channel with transparent walls is $47 \times 47$ cm$^2$, a diameter of the jet nozzle is  $D = 10$ mm and the diameter of the jet in the probed region is about 35 mm. Consequently, the side walls of the channel  weakly affect the development of the jet. The
optics (4) produces a light sheet (5), and the scattered light from
the probed flow region (6) is recorded with the CCD camera (7), see Fig.~\ref{Fig1}.
The laser sheet thickness is 1 mm. Note that in the present paper we study mixing of particles at the external boundary of a jet and we do not investigate the dynamics of jet.

Flow measurements have been conducted with digital Particle Image
Velocimetry (PIV) using a LaVision Flow Master III system to determine the jet velocity field (see, e.g., \cite{AD91,RWK98,W00}). The light sheet is provided by a double PIV Nd-YAG laser, Continuum Surelite $2 \times 170$ mJ. The beam of the laser is directed with the aid of a few prisms to the light sheet optics
that comprises spherical and cylindrical Galilei telescopes with
tunable divergence and adjustable focus length. A progressive-scan
12 Bit digital CCD camera with dual-frame-technique for cross
correlation captures the images of $1280 \times 1024$ pixels with a size $6.7 \, \mu$m $\times 6.7 \, \mu$m.
A programmable Timing Unit (an interface card of the system PC)
generates sequences of pulses to control the laser, the camera and
data acquisition. The software package DaVis 7 is applied to
control all hardware components and for image acquisition and
visualization. Jet flow images are analyzed with the package
supplemented with the software developed in this study.

In this experimental study we use methods of measurements applied previously in different studies of turbulent transport of particles (for details see \cite{BEE04,EEKR06B,EEKR06C}). In particular, we use PIV and an Image Processing Technique based on the analysis of the intensity of the Mie scattering to determine the spatial distribution of tracer particles (see, e.g., \cite{G01}). As a tracer we use smoke produced by incense particles sublimation. The submerged air jet, seeded with the incense smoke particles
of sub-micron sizes with the mean diameter $0.7 \mu$m, flows into the still air inside the chamber. An estimate of the mean response time of these particles $\tau_p$ to a step change of the velocity is of the order of $10^{-6}$ s. Since the latter value is much smaller than the Kolmogorov time
$\tau_{_{K}}$ of the flow in the jet ($\tau_{_{K}}$ varies in the rage from $10^{-4}$ s to $10^{-3}$ s), and smoke particles Stokes number St$ = \tau_p/ \tau_{_{K}} \ll 1$, it can be safely assumed that these particle follow the flow faithfully.

Laser sheet light scattered by the incense smoke particles
is captured with a CCD camera. Series of 50 pairs of images with the
adjustable time delay are acquired with a frequency of 2 Hz and are
stored for ensemble and spatial averaging of flow characteristics.
We measure spatial particle distribution in a flow area of
$18.4 \times 18.4$  cm$^2$ with a spatial resolution of
$1024 \times 1024$  pixels that is 0.18 mm/pixel.

In the experiments the jet fluid is characterized
by a high concentration of the tracer particles, and the external
fluid is characterized by a low concentration of the particles.
Every recorded image is normalized by a light intensity measured just
at the jet entrance into the chamber in order to eliminate the effects associated with a change of concentration of the incense smoke. We
measure parameters of mixing in a local system of coordinates $(y,z)$, where $y$-axis is directed along the averaged jet-ambient fluid interface, $z$-axis is perpendicular to this interface (see Fig.~\ref{Fig1}). The jet parameters are
determined in intervals L, where the flow characteristics are
statistically homogeneous.

\section{Data processing based on the measured phase function}

In this section we use the approach applied previously
in the analysis of Rayleigh-Taylor instability (see \cite{H06}).
This approach is based on the measured phase function,
$X(y, \phi)$, defined as follows: $X(y,\phi)=1$
when the point $(y, \phi)$ is within the jet fluid and $X(y, \phi)=0$
elsewhere  (see \cite{D96}), where $\phi$ is the polar angle
(see Fig.~\ref{Fig1}).  We analyze the statistical properties of the system by averaging over an interval of $L$ along the $y$ direction at fixed angles $\phi$, i.e., $ \langle X \rangle= L^{-1} \int_0^L X(y, \phi) \,dy $ , where $L$ is the length of the interval in $y$ direction (see Fig.~\ref{Fig1}).

\begin{figure}
\vspace*{2mm} \centering
\includegraphics[width=6cm]{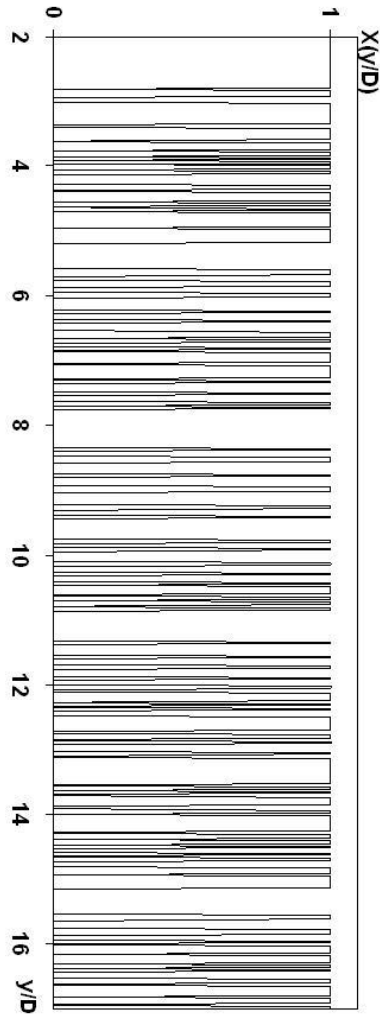}
\caption{\label{Fig2} The measured phase function,
$X(y)$ at fixed angle $\phi$. Here $\Delta$ are the sizes of regions with $X(y)=1$ that are occupied by the jet fluid.}
\end{figure}

Let us introduce the first and the second moments of the measured
phase function: $\alpha(\phi) = \langle X \rangle$ and $\varsigma(\Delta, \phi) = \langle X(y -\Delta/2, \phi) \, X(y +\Delta/2, \phi) \rangle$ , where $\alpha(\phi)$ characterizes the fraction of $y$ which is occupied by the jet fluid, $\Delta$ are the sizes of regions (with $X(y)=1)$ occupied by the jet fluid (see Fig.~\ref{Fig2}). The study of the mixed state of two fluids induced by Rayleigh-Taylor instability conducted in \cite{H06} indicates that PDF of the random size $\Delta$ of the regions occupied by the jet fluid can be described by Gamma distribution
\begin{eqnarray}
f(\Delta) = {1 \over \Gamma(2-r)} \, {a \, L \over \lambda^2}
\, {\exp(-\Delta/\lambda) \over (\Delta/\lambda)^r}  \;,
 \label{A1}
\end{eqnarray}
where $\Delta$ are the sizes of the regions occupied by the jet fluid, the scale $\lambda$ and the exponent $r$ are parameters characterizing the length scale and a deviation from the exponential PDF. A ratio of the moments
\begin{eqnarray}
\lambda_n = {\int_0^\infty \Delta^{n+r} \,f(\Delta) \, d\Delta \over \int_0^\infty \Delta^{n+r-1} \,f(\Delta) \, d\Delta}  \;,
 \label{A2}
\end{eqnarray}
determines a characteristic scale $\lambda_n$ , where $n$ is the number of the statistical moment. Equation~(\ref{A1}) for PDF implies that
\begin{eqnarray}
{\lambda_n \over n} =const \; .
 \label{A3}
\end{eqnarray}
This property of the PDF of the sizes $\Delta$ of the regions occupied by the jet fluid, is valid if the parameter $\Delta$ in Eqs.~(\ref{A1}) and ~(\ref{A2}) varies in the infinite interval $0<\Delta< \infty$.

\begin{figure}
\vspace*{2mm} \centering
\includegraphics[width=6cm,angle=90]{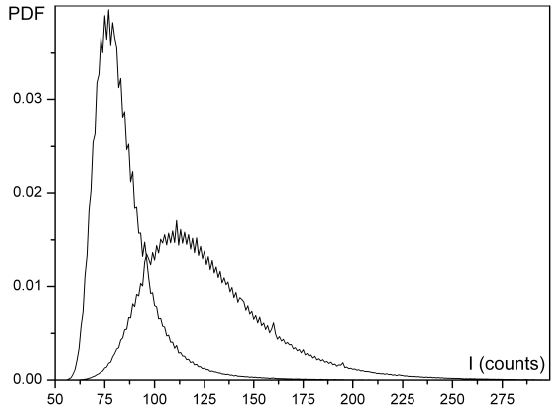}
\caption{\label{Fig3} Measured PDF of light intensities in the jet (right line) and in the ambient fluid (left line).}
\end{figure}

We show below that the measured PDF of the phase function in our experiments can be reasonably approximated by Gamma distribution.
In the present analysis we use images of a jet obtained in our
experiments, whereby the interface between a jet and an ambient fluid is determined for a mean jet image averaged over 50 independent instantaneous images. The images are converted into a binary form using the following procedure. The threshold for the image transformation into a binary form is chosen equal to the intensity whereby the PDF of the scattered light intensity in the jet equals to that in the ambient fluid (see Fig.~\ref{Fig3}, where this threshold is 96 counts). Thereafter all pixels of every image where the scattered light intensity is above
the threshold value, are assigned the value of the phase function $X(y, \phi)$ equal 1 and are categorized  as a jet fluid, while pixels with light intensity below the threshold are assigned the value of the phase function equal 0 and are categorized as an ambient fluid.

After implementing this procedure the mean image of the
instantaneous binary jet images is obtained. The mean image is
transformed into a binary form with the threshold 0.5. In particular,
the pixels, where a probability of a jet fluid appearance was higher
then 0.5 were assigned the value of the phase function equal 1,
while all the rest were assigned the value of the phase function
equal 0. The boundary of a jet is a line, where a mean value of a
phase function equals 0.5. The threshold adopted in this study
(the intensity whereby the PDFs of the scattered light intensity in the jet equals to that in the ambient fluid), allows to decrease a systematic error during the transformation into a binary form. Our analysis has shown that the obtained results on mixing of particles at the external boundary of a jet are not very sensitive to a particular choice of the threshold.

We can roughly distinguish between three regions
in the jet down steam the nozzle. In the first region having the size
$\sim D$ the jet is stable, where $D$ is a diameter of a nozzle.
The second region includes an unstable part of the jet, whereby the
Kelvin-Helmholtz instability might be the main feature of the flow,
and this region extends up to a distance of about $\sim 3 D$. The third region down steam is a turbulent flow region that is the subject of our study.

\begin{figure}
\vspace*{2mm} \centering
\includegraphics[width=8cm]{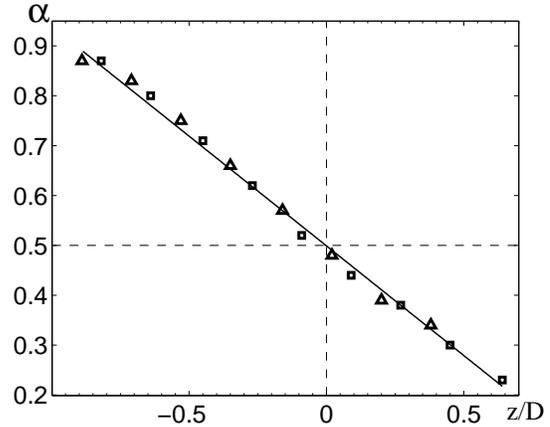}
\caption{\label{Fig4} First moment $\alpha$ of the phase function for different $z$ inside and outside the jet determined in the experiments with different global Reynolds numbers: ${\cal R}=8.4 \times 10^3$ (triangles) and ${\cal R}=10^4$ (squares). Here $z=0$ corresponds to the mean boundary of the jet, $z<0$ inside the jet and $z>0$ outside the jet.}
\end{figure}

\begin{figure}
\vspace*{2mm} \centering
\includegraphics[width=8cm]{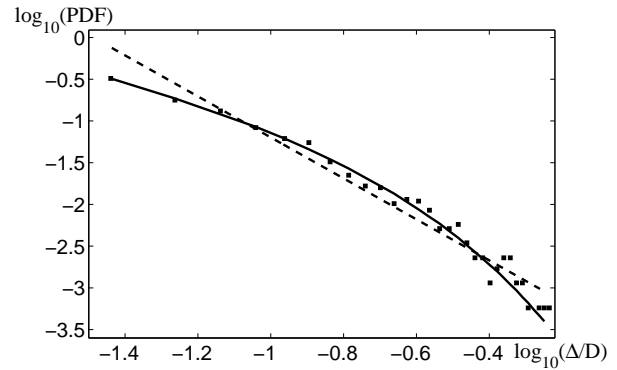}
\caption{\label{Fig5} PDF of the phase function of the jet fluid penetrating into an ambient flow (squares) and its approximation (solid line) for $z=0$  by the Gamma distribution (see Eq.~(\ref{A1})) with the following parameters: $r =1.03$, $\lambda/D = 0.13$, and $A = 0.182$. For comparison, a power-law fitting curve $A_1 \, \xi^{-r_1}$  (dashed line) is also shown, where $r_1 =2.46$, $\lambda/D = 0.13$, and $A_1 = 3.36 \times 10^{-2}$.}
\end{figure}

A phase function, $X(y,z)$, is determined for each image in the interval of $y/D$ from $4.5$ to $10.9$ along the different directions with a given $\phi$ (or for different values of $z$ in the range  $\pm D$), see Fig.~\ref{Fig1}. We determine a mean value, $\alpha$, of the phase function $X(y,z)$ averaged over the ensemble of the jet images. For instance, the function $\alpha(z)$ is shown in Fig.~\ref{Fig4} for two experiments with different global Reynolds numbers, ${\cal R}=8.4 \times 10^3$ and ${\cal R}=10^4$. Here ${\cal R} = V D /\nu$ is the global Reynolds number based on the mean jet velocity $V$ at the nozzle exit and the diameter $D$ of a jet nozzle. Although we performed experiments for two values of the Reynolds number, we did not study the dependence of mixing on the Reynolds number.
The line $z=0$ corresponds to the mean boundary of the jet, $z<0$ inside the jet and $z>0$ outside the jet. The mean value $\alpha$ of the phase function decreases along the positive direction of the axis $z$.

In our analysis of mixing of particles at the external boundary of a jet it is not important whether the jet at distances $y/D =4.5-10.9$ could be affected by the initial conditions. Indeed, the interface between the jet and the ambient fluid is determined for a mean jet image averaged over 50 independent images that are acquired with a frequency of 2 Hz. The successive pairs of images are not correlated since the turbulence correlation time  $\tau_0$ is of the order of $10^{-3}$ s and the characteristic time of the mean flow in the image scale is of the order of $10^{-2}$ s. The estimate for the turbulence correlation time is obtained using the measured characteristic turbulent velocity $u_0 \sim (1-2) \times 10^3$ cm/s, the maximum scale of turbulent motions $l_0 \sim 1$ cm and $\tau_0 = l_0/u_0 \sim 10^{-3}$ s.

PDF of the phase function of a jet fluid penetrating into an ambient fluid is determined for the same intervals of $y/D$ for different $z$ using parametric histogram estimator (see, e.g., \cite{SI86}). The obtained PDF is approximated by the Gamma distribution~(\ref{A1}) using three adjustable parameters $r$, $\lambda$ and $A=a \, L/ \Gamma(2-r)$. The measured in our experiments PDF of the phase function and its approximation by the Gamma distribution, are shown in Fig.~\ref{Fig5}, where the parameters of the Gamma distribution are $r =1.03$, $\lambda/D = 0.13$, and $A = 0.182$. These PDFs vary  more slowly for large values of $\Delta$ than for small $\Delta$. For comparison, a power-law fitting curve $A_1 \, \xi^{-r_1}$  is also shown in Fig.~\ref{Fig5}, where $r_1 =2.46$, $\lambda/D = 0.13$, and $A_1 = 3.36 \times 10^{-2}$. Clearly, the measured PDF cannot be fitted by a power-law function (see Fig.~\ref{Fig5}).

\begin{figure}
\vspace*{2mm} \centering
\includegraphics[width=8cm]{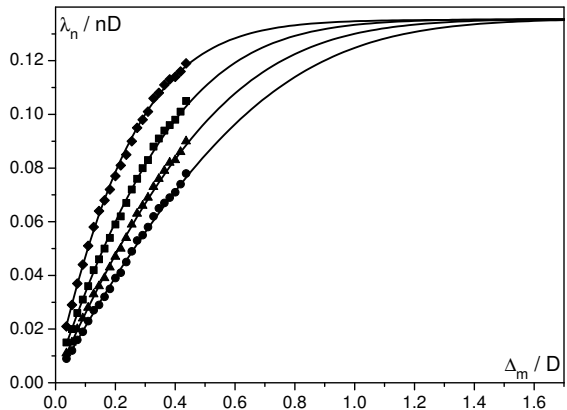}
\caption{\label{Fig6} Ratios of the moments $\lambda_n/nD$ of the measured PDF versus the length $\Delta_m/ D$  for $n$ varied from 1 (squares) to 4 (circles), where $\Delta_m$ is the maximum size of the observed jet fluid inclusions. Ratios $\lambda_n/nD$  are also determined using a direct fit of the measured PDF to Gamma distribution [Eqs.~(\ref{A1})-(\ref{A2})] with the following parameters: $r =1.03$, $\lambda/D = 0.13$, and $A = 0.182$, where $n=1$ is the upper solid curve, and $n=4$ is the lower solid curve).}
\end{figure}

\begin{figure}
\vspace*{2mm} \centering
\includegraphics[width=8cm]{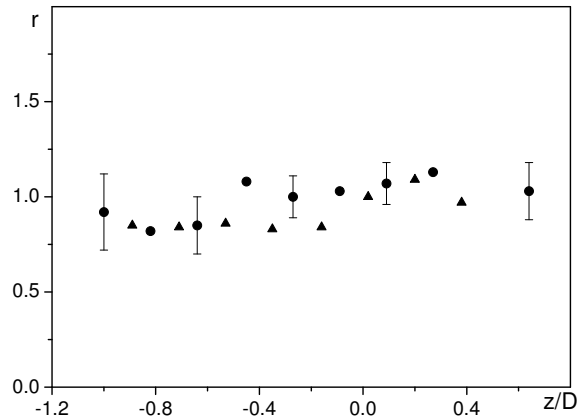}
\caption{\label{Fig7} Dependence of the exponent $r$ on the distance from the jet boundary for ${\cal R}=8.4 \times 10^3$ (triangles) and ${\cal R}=10^4$ (circles).}
\end{figure}

\begin{figure}
\vspace*{2mm} \centering
\includegraphics[width=8cm]{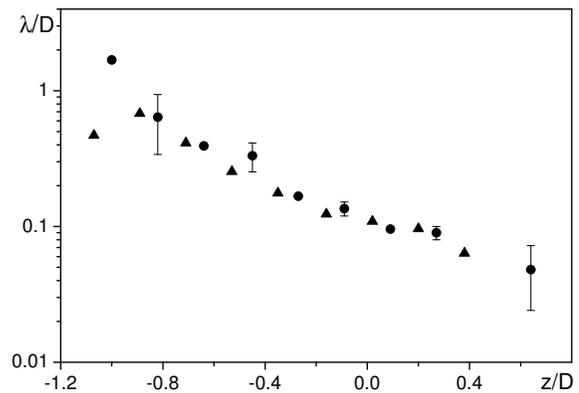}
\caption{\label{Fig8} Dependence of a scale $\lambda$ on a distance $z/D$ from the jet boundary for ${\cal R}=8.4 \times 10^3$ (triangles) and ${\cal R}=10^4$ (circles).}
\end{figure}

The ratios of the moments $\lambda_n/n$ for the measured PDF versus the length $\Delta_m$ are shown in Fig.~\ref{Fig6}, where $\Delta_m$ is the maximum size of the observed jet fluid inclusions. A range of our experimental data is limited to $\Delta_m/ D  = 0.6$, where the probability of the event with $\Delta/ D = 0.6$ is about $3 \times 10^{-4}$. Consequently, it can be expected that only one such event occurs in a sample with the size of about 3300 events (not to mention the events with $\Delta/ D > 0.6$). Therefore, it is not feasible to expand significantly the $\Delta_m/ D$ range of the observed events, since a probability of the events with $\Delta/ D > 0.6$ is extremely low. Using Eq.~(\ref{A2}) we also determine the scale $\lambda_n/n$ for $n$ varied from 1 to 4 using a direct fit (solid lines in Fig.~\ref{Fig6}) of the measured PDF to Gamma distribution with the following parameters: $r =1.03$, $\lambda/D = 0.13$, and $A = 0.182$.
Inspection of Fig.~\ref{Fig6} shows that if the range of integration $\Delta_m$ is too short (i.e., $\Delta_m < 10 \, \lambda_n/n$), the ratios $\lambda_n/n$ are different for different $n$. On the other hand, when $\Delta_m > 10 \, \lambda_n/n$, the ratios $\lambda_n/n$ are independent of $n$. Therefore, the method of determining the parameters of PDF [see Eq.~(\ref{A1})], using the property~(\ref{A3}) of Gamma distribution, $\lambda_n/n$=const, is not appropriate if $\Delta_m < 10 \, \lambda_n/n$. The latter condition corresponds to all our experiments (see Fig.~\ref{Fig6}). This implies that in our experiments we cannot account for large $\Delta$ when we determine $\lambda_n$. Therefore, the method based on Eq.~(\ref{A3}) can be used only when $\Delta_m > 10 \, \lambda_n/n$.

The parameters $r$ and $\lambda$ of the PDF of the phase function versus $z$ determined in the experiments are shown in Figs.~\ref{Fig7} and~\ref{Fig8}. The characteristic scale $\lambda$ that is shown in a logarithmic scale (see Fig.~\ref{Fig8}), varies significantly, decreasing by 1.5 orders of magnitude from the internal to the external regions of a jet. This dependence is close to an exponential function. The behaviour of the functions $\lambda(z)$ and $\alpha(z)$ is similar (see Figs.~\ref{Fig4} and~\ref{Fig8}). Indeed, $\alpha$ is the first moment (mean value) of the phase function, while $\lambda$ should coincide with the first moment $\lambda_1$ if the sizes $\Delta$ of the regions occupied by the jet fluid vary from 0 to very large values. For negative $z$ (i.e., inside the jet) the size $\lambda$ of the region occupied by the jet fluid is much larger than that for positive $z$ (i.e., outside the jet), whereby the concentration of the jet fluid is small.

It must be noted that we perform experiments in the region located at the distance $4.5 - 11$ cm downstream from the nozzle. The reason is that in this region the gradient of the mean particle number density at the external boundary of a submerged turbulent jet is very large. In this region the tangling mechanism that is responsible for particle mixing, is more pronounced. The tangling mechanism  depends on the gradient of the mean particle number density (see Sect.~IV). Increase of the distance from the nozzle, is accompanied by a decrease of the gradient of the mean particle number density. In addition, in the regions located far from the nozzle, the effect of the bottom wall of the chamber (e.g., appearance of the secondary flows) becomes stronger.

Note that mixing in water jet flows was also studied in recent experiments \cite{VD03} (see also \cite{V04,V08}). In these experiments the water jet with ink having a diameter $D=0.8$ cm and the global Reynolds numbers ${\cal R} \sim 10^4$, discharges into a square duct with a width $L=3$ cm. The measurements described in \cite{VD03} were performed in the duct section at the distance $x/D$ in the range from 6.6 to 22.5 downstream of the jet entrance (see Figure 2b in \cite{VD03}). Inspection of Figure 1 in \cite{VD03} reveals that the mixed part of the flow spreads over the whole cross-section of the channel after $\sim 4.5$ cm ($x/D \sim 5.6$) downstream of the jet entrance. Consequently, most of the measurements in \cite{VD03} were performed in the non-turbulent duct flow with the global Reynolds number ${\cal R} \sim 2 \times 10^3$ (based on the width of the duct and the mean flow velocity). The flow in another experiment reported in \cite{VD03} is also laminar (${\cal R}\sim 0.1$). In these experiments it was found that the PDF of the concentration of the dye in the channel can be described by Gamma distribution with a positive parameter $r=n-1 \sim 3 - 84$  in a power function. In our study we consider a kinematic problem of mixing at the turbulent fluid-ambient fluid interface and found that the PDF of the random size of the regions occupied by the jet fluid can be described by a Gamma function with a negative parameter $r \sim
- 0.9$ in a power function. These findings imply the absence of power-law scaling.

Although the PDF of phase function was fitted in Fig.~\ref{Fig5} by Gamma function, one of the main findings of this section is that the PDF of phase function does not have a universal scaling and cannot be described by a power-law function.

\section{Theoretical modelling of turbulent mixing and comparison with experimental results}

In the previous section we analyze experimental data using the measured phase function. In this section we use another approach in the analysis of turbulent mixing of particles at the external boundary of the submerged turbulent jet. This approach is based on the analysis of the two-point second-order correlation function of the particle number density fluctuations generated by tangling of the gradient of the mean particle number density by the turbulent velocity field. This gradient is formed at the external boundary of a submerged turbulent jet.

The equation for the number density $n(t,{\bf x})$ of particles advected by a fluid velocity field reads:
\begin{eqnarray}
\frac{\partial n}{\partial t} + {\bf \nabla \cdot} (n \,{\bf v}) =
D_m\, \Delta n \;,
 \label{B1}
\end{eqnarray}
where $D_m$ is the coefficient of molecular (Brownian) diffusion, ${\bf v}(t,{\bf x}) = {\bf V}+ {\bf u}$ is the particle velocity field, ${\bf V} = \langle {\bf v} \rangle$ is the mean velocity and ${\bf u}$ are velocity fluctuations. Equation~(\ref{B1}) implies conservation of the total number of particles in a closed volume. In this study we consider incompressible particle turbulent velocity field $({\bf \nabla \cdot} \,{\bf u} = 0)$. This implies that we neglect a small compressibility of particle velocity field caused by particle inertia, because the particle size in our experiments is less that $1 \mu$m. We also neglect very small compressibility of the air velocity. Consider $\Theta(t,{\bf x}) = n - N$, the deviation of $n$ from the mean number density of particles $N$. The equation for fluctuations of the particle number density $\Theta(t,{\bf x})$ reads:
\begin{eqnarray}
\frac{\partial \Theta}{\partial t} + {\bf \nabla \cdot} (\Theta \,{\bf u}- \langle \Theta \,{\bf u}\rangle) = D_m\, \Delta \Theta - ({\bf u} {\bf \cdot \nabla}) N \;,
 \label{RB1}
\end{eqnarray}
(see, e.g., \cite{BA71,BA59}). Using Eq.~(\ref{RB1}) we derive equation for the evolution of the two-point second-order correlation function of the particle number density, $\Phi(t,{\bf R}) \equiv \langle \Theta(t,{\bf x}) \Theta(t,{\bf x}+{\bf R}) \rangle$:
\begin{eqnarray}
{\partial \Phi \over \partial t}  = 2 \Big[D_m \delta_{ij} + D_{ij}^{^{T}}(0) - D_{ij}^{^{T}}({\bf R})\Big] \, \nabla_{i} \nabla_{j} \Phi + I \;,
 \label{B2}
\end{eqnarray}
where $D_{ij}^{^{T}} ({\bf R})$ is the turbulent diffusion tensor (see below), $\delta_{ij}$ is the Kronecker tensor, $I = D_{_{T}} \, (\bec {\nabla} N)^2 \exp(-c_\ast R/l_0)$ is the source of particle number density fluctuations at the jet-ambient fluid interface, $D_{_{T}} \propto l_0 \,u_0$ is the turbulent diffusion coefficient, $u_0$ is the characteristic turbulent velocity in the maximum scale $l_0$ of turbulent motions, $c_\ast > 1$ is a free constant and the angular brackets denote ensemble averaging. The source function $I$ is related to the last term in the right hand side of Eq.~(\ref{RB1}).

When $\bec {\nabla} N \not= 0$ the nonzero source $I$ results in generation of fluctuations of the particle number density caused by tangling of the gradient of the mean particle number density by the turbulent velocity field. In this section we study this effect in detail. Equation~(\ref{B2}) has been derived in \cite{KR68} for a delta-correlated in time random velocity field. For a turbulent velocity field with a finite correlation time Eq.~(\ref{B2}) has been derived in \cite{EKR02} by means of stochastic calculus, i.e., Wiener path integral representation of the solution of the Cauchy problem for Eq.~(\ref{B1}), using Feynman-Kac formula and Cameron-Martin-Girsanov theorem. The comprehensive description of this approach can be found in \cite{EKR95,EKR02,ZRS90}.

The turbulent diffusion tensor, $D_{ij}^{^{T}} ({\bf R})$, is given by the following formula:
\begin{eqnarray}
D_{ij}^{^{T}} = \int_{0}^{\infty} \langle u_{i}\big[0,\bec{\xi}(t,{\bf x}|0)\big]
\,u_{j}\big[\tau,\bec{\xi}(t,{\bf x}+{\bf R}|\tau)\big] \rangle \,d \tau \;,
\nonumber\\
 \label{B3}
\end{eqnarray}
(see \cite{EKR02}). Here the Wiener trajectory $\bec{\xi}(t,{\bf x}|s)$ (which is often called the Wiener path) is defined as follows:
\begin{eqnarray}
\bec{\xi}(t,{\bf x}|s) &=&{\bf x} - \int^{t}_{s} {\bf u}
[\tau,\bec{\xi}(t,{\bf x}|\tau)] \,\,d \tau - \sqrt{2 D_m} \, {\bf
w}(t-s) ,
\nonumber\\
 \label{B33}
\end{eqnarray}
where ${\bf w}(t)$ is the Wiener random process which describes the
Brownian motion (molecular diffusion). The Wiener random process
${\bf w}(t)$ is defined by the following properties: $\langle {\bf
w}(t) \rangle_{\bf w}=0\,, $ $\, \langle w_i(t+\tau) w_j(t)
\rangle_{\bf w}= \tau \delta _{ij}$, and $ \langle \dots
\rangle_{\bf w} $ denotes the mathematical expectation over the
statistics of the Wiener process. The velocity $u_i[\tau,
\bec{\xi}(t,{\bf x}|\tau)]$ describes the Eulerian velocity
calculated at the Wiener trajectory. For a random incompressible velocity field with a finite correlation time the tensor of turbulent diffusion is given by $D^{^{\rm T}}_{ij} ({\bf R}) \approx \tau^{-1} \langle \xi_i(t,{\bf x}|0) \, \xi_j(t,{\bf x} +{\bf R}|\tau) \rangle $ (see \cite{EKR02}). Hereafter $\langle ... \rangle$ denotes averaging over the statistics of both, turbulent velocity field and the Wiener process.

If the turbulent velocity field is not delta-correlated in time (e.g., the correlation time is small yet finite), the tensor of turbulent diffusion, $D_{ij}^{^{T}} ({\bf R})$, is compressible, i.e.,
$(\partial / \partial R_i) D_{ij}^{^{T}} ({\bf R}) \not= 0$ (for details see \cite{EKR02}). Let us consider the parameter $\sigma_{_{T}}$ that characterizes the degree of compressibility of the tensor of turbulent diffusion:
\begin{eqnarray}
\sigma_{_{T}} \equiv \frac{\nabla_i \nabla_j D^{^{\rm T}}_{ij}({\bf R})}{\nabla_i\nabla_j D^{^{\rm T}} _{mn}({\bf R}) \epsilon_{imp}\epsilon_{jnp} } \approx {\langle (\bec{\nabla} {\bf \cdot} \bec{\xi})^{2} \rangle \over \langle (\bec{\nabla} {\bf \times} \bec{\xi})^{2} \rangle} \;,
 \label{S1}
\end{eqnarray}
where $\epsilon_{ijk}$ is the fully antisymmetric Levi-Civita unit tensor. Note that when the turbulent velocity field is a delta-correlated in time random process, $\bec{\nabla} {\bf \cdot} \bec{\xi} \propto \bec{\nabla} {\bf \cdot} {\bf u}$, while for a turbulent flow with the finite correlation time $\bec{\nabla} {\bf \cdot} \bec{\xi}$ has a contribution that is independent of $\bec{\nabla} {\bf \cdot} {\bf u}$.

Let us discuss the assumptions underlying the above model of particle transport in turbulent flow. We use the tensor of turbulent diffusion $D^{^{\rm T}}_{ij} ({\bf R})$ for isotropic and homogeneous turbulent flow. In our experiments the velocity field is generally anisotropic. We determine the two-point second-order correlation function $\Phi(t,{\bf R})$ in the vicinity of the mean boundary of the jet along $y$-axis, so that the size of the probed region in $z$ direction is small. It is of the order of the maximum scale of turbulent motions $l_0$.

In addition, a main contribution to the level of fluctuations of particle number density is due to the mode with the minimum damping rate (see Eqs.~(\ref{T31}) and~(\ref{T32}) below). This mode is an isotropic solution of Eq.~(\ref{B2}). Consequently, it is plausible to neglect the anisotropic effects. This assumption is also supported by our measurements of the two-point second-order correlation function $\Phi(t,{\bf R})$ determined along $y$-axis for different values of $z$ inside and outside the jet (see Fig.~\ref{Fig13} below). The difference between these correlation functions determined for different $z$ is very small. We perform measurements in the range along $y$-axis whereby the turbulence is nearly homogeneous. On the other hand, the turbulence parameters $l_0$ and $u_0$ vary slowly with $z$ in the probed region.

Note that the mechanism of mixing related to the tangling of the gradient of the mean particle number density at the external boundary of a submerged turbulent jet by the turbulent velocity field, is enough robust. The properties of the tangling are not very sensitive to the exponent of the energy spectrum of the background turbulence. The requirements that turbulence should be an isotropic, homogeneous, and should have a very long inertial range (a fully developed turbulence), are not necessary for the tangling mechanism.

Anisotropy effects can complicate the theoretical analysis, but do not introduce new physics in the mixing process.
The reason is that the main contribution to the the tangling mechanism is at Kolmogorov (viscous) scale of turbulent motions. At this scale turbulence can be considered as nearly isotropic, while anisotropy effects can be essential in the vicinity of the maximum scale of turbulent motions.

Using these arguments, we consider the tensor $D^{^{\rm T}}_{ij} ({\bf R})$ in the following form:
\begin{eqnarray}
D^{^{\rm T}}_{ij} ({\bf R}) &=& D_{_{\rm T}} \, \Big[ [F(R) + F_c(R)] \delta_{ij} + R F'_c \, {R_i R_j \over R^2}
\nonumber\\
&& + {R F' \over 2} \Big(\delta_{ij} - {R_i R_j \over R^2} \Big) \Big]
\;,
\label{T15}
\end{eqnarray}
$D_{_{T}} = l_0 \,u_0/3$ is the turbulent diffusion coefficient, $F(0) = 1 - F_c(0)$ and $F'=dF/dR$. The function $F_c(R)$ describes the compressible (potential) component, whereas $F(R)$ corresponds to vortical (incompressible) part of the turbulent diffusion tensor. Using Eqs.~(\ref{B2}) and~(\ref{T15}) we derive equation for the two-point second-order correlation function $\Phi(R)$ written in a dimensionless form:
\begin{eqnarray}
{\partial \Phi \over \partial t}  = {1 \over M(R)} \Big[\Phi'' + 2 \, \Big({1 \over R} + \chi(R) \Big) \, \Phi' \Big] + I \;,
\nonumber\\
\label{T1}
\end{eqnarray}
where time $t$ is measured in units of $\tau_0=l_0/u_0$, distance
$R$ is measured in units of $l_0$, and
\begin{eqnarray}
{1 \over M(R)} &=& {2 \over {\rm Pe}} + {2 \over 3} [1 - F - (R F_c)']\;,
\label{L1}\\
\chi(R) &=&  - {M(R) \over 3} (F - 2 F_c)' \;,
\label{L2}
\end{eqnarray}
and ${\rm Pe} = u_0 l_0 / D_m$ is the Peclet number. The two-point correlation function $\Phi(R)$ satisfies the following boundary conditions: $\Phi(R=0) = 1$ and $\Phi(R \to \infty) = 0$. This function
has a global maximum at $R=0$ and therefore it satisfies the conditions:
\begin{eqnarray*}
\Phi'(R=0) &=& 0\,, \quad \Phi''(R=0) < 0 \;,
\\
\Phi(R=0) &>&  |\Phi (R>0)| \; .
\end{eqnarray*}

Let us introduce a function
\begin{eqnarray}
\psi(t,{\bf R}) = R \, \Phi(t,{\bf R}) \exp \biggl[ \int_0^R \chi(x) \,dx
\biggr] \; .
\label{T20}
\end{eqnarray}
Equations~(\ref{T1}) and~(\ref{T20}) yield the following equation for the function $\psi(R)$:
\begin{eqnarray}
{\partial \psi \over \partial t} = {1 \over M(R)} {\partial^2 \psi \over \partial R^2} - U(R) \psi + I_0 \;,
\label{T26}
\end{eqnarray}
where
\begin{eqnarray}
U(R) &=& {\chi(R) \over M(R)} \, \biggl( {2 \over R} + \chi(R) + {\chi' \over
\chi} \biggr) \;,
\label{L3}\\
I_0 &=& R \, I(R) \, \exp \biggl[ \int_0^R \chi(x) \,dx
\biggr] \; .
\label{L4}
\end{eqnarray}
In the analysis we use a quantum mechanics analogy, whereby Eq.~(\ref{T26}) is regarded as a one-dimensional Schr\"{o}dinger equation with a variable mass. We seek for the solution of Eq.~(\ref{T26}) in the following form:
\begin{eqnarray}
\psi(t,R) = \sum_{p=1}^{\infty} \phi_p(t) \Psi_p(R) \;,
\label{T27}
\end{eqnarray}
where $ \Psi_p(R) $ are the eigenfunctions determined by the following equation:
\begin{eqnarray}
\frac{1}{M(R)} {d^2 \Psi_p \over dR^2} - [ 2 \gamma_p + U(R)] \Psi_p = 0
\; .
\label{Q4}
\end{eqnarray}
The condition of the orthogonality for the
eigenfunctions reads:
\begin{eqnarray}
\int_0^{\infty} M(R) \Psi_i(R) \Psi_j(R) \,dR = \delta_{ij} \; .
\label{T28}
\end{eqnarray}
Substituting solution~(\ref{T27}) into Eq.~(\ref{T26}), multiplying the obtained equation by $ \Psi_l(R) $ and integrating over $R$ yields the following equation for the function $\phi_p(t)$:
\begin{eqnarray}
{d  \over d t} \phi_p - \gamma_p \phi_p = {2 \over 3} l_0^2
(\bec {\nabla} N)^2 \; .
\label{T29}
\end{eqnarray}
In derivation of Eq.~(\ref{T29}) we use Eq.~(\ref{T28}) and take into account that
\begin{eqnarray*}
&& \int_0^{\infty} M(R) \,R \,I(R) \,\exp \biggl(
\int_0^R \chi(x) \,dx \biggr) \Psi_p(R) \,dR
\nonumber\\
&& \simeq
{2 \over 3 c_\ast} l_0^2 (\bec{\nabla} N)^2
\;,
\end{eqnarray*}
where the function $M(R) I(R)$ is strongly localized and is approximated by the delta function. We also assume here that the basis of the eigenfunctions $\Psi_p(R)$ is complete. The solution of Eq.~(\ref{T29}) for the function $\phi_p(t)$ with the initial condition $\phi_p(t=0) =0$ reads:
\begin{eqnarray}
\phi_p(t) = {2 \over 3 c_\ast} {l_0^2 \over \vert \gamma_p \vert} (\bec{\nabla} N)^2 \Big[1 - \exp (- \vert \gamma_p \vert t) \Big] \; .
\label{T31}
\end{eqnarray}
Equations~(\ref{T20}), (\ref{T27}) and~(\ref{T31}) yield the correlation function $\Phi(t,R)$:
\begin{eqnarray}
\Phi(t,R) &=& {2 \over 3 c_\ast} \biggl({l_0 \over L_\ast} \biggr)^2
(\delta N)^2 \sum_{p=1}^{\infty} {\Phi_p(R) \over \vert \gamma_p
\vert}
\nonumber\\
&& \times \Big[1 - \exp (- \vert \gamma_p \vert t)\Big]  \;,
\label{T32}
\end{eqnarray}
where we used an estimate $ (\bec {\nabla} N)^2 \sim (\delta N)^2
/ L_\ast^2 .$  The main contribution to the correlation function $
\Phi(t,R) $ for $ t \gg \vert \gamma_p \vert ^{-1} $ in Eq.~(\ref{T32}) is due to the mode with the minimum damping rate $\vert \gamma_p \vert $, i.e., for the mode with $p=1$.

Particular formulas for the potential $U(R)$ and the mass $M(R)$ depend on the functions $F(R)$ and $F_c(R)$. For instance, we may choose these functions in the following form
\begin{eqnarray}
F(R) &=& {1 \over 1 +\sigma_{_{T}}} \, \exp[- f(R)] \;,
\label{C10}\\
F_c(R) &=& {\sigma_{_{T}} \over 1 +\sigma_{_{T}}} \, \exp[- a_c \, f(R)] \;,
\label{C11}\\
f(R) &=& {R^2 \over {\rm Re}^{-1/2} + R^{2/3}}  \;,
\label{C12}
\end{eqnarray}
where ${\rm Re}= u_0 l_0 / \nu$ is the Reynolds number based on turbulent velocity $u_0$ and maximum scale $l_0$ of turbulent motion. Equation~(\ref{C12}) is similar to an interpolation formula derived by Batchelor for the correlation function of the velocity field that is valid for a turbulence with Kolmogorov spectrum in the inertial range and for random motions in the viscous range of scales (see, e.g., \cite{MY75,Mc90}). In particular, in the inertial range of turbulent scales, ${\rm Re}^{-3/4} \ll R \ll 1$, the correlation function for the turbulent diffusion tensor is $F(R) \propto 1 - R^{4/3}$, where $R$ is measured in units of maximum scale of turbulent motions $l_0$. The corresponding correlation function for the turbulent velocity field $\propto 1 - R^{2/3}$. The difference between the scalings of the turbulent diffusion tensor and the correlation function of the turbulent velocity field is caused by the scaling of correlation time $\tau(R) \propto R^{2/3}$. On the other hand, in the viscous range, $R \ll {\rm Re}^{-3/4}$, the correlation function for the turbulent diffusion  tensor is $F(R) \propto 1 - {\rm Re}^{1/2} R^{2}$. This correlation function is similar to that for the velocity field because in the viscous range the correlation time is independent of scale. On the other hand, for large scales $R \gg 1$ there is no turbulence, so that for $R > 1$ the functions  $F(R)$ and $F_c(R)$ should decrease sharply to zero.

The particular choice of the functions $F(R)$, $\, F_c(R)$ and $f(R)$ in the paper describes the well-known properties of turbulent velocity field obtained from laboratory experiments, numerical simulations and theoretical studies (see, e.g., \cite{MY75,Mc90}). In this study we choose the exponential form for the functions $F(R)$ and $F_c(R)$. The final results are not sensitive to the form for these functions at large scales $R \gg 1$. Note also that the particular choice of the function $f(R)$ is also not very  important. It should include turbulent motions in the inertial range (e.g., a turbulence with Kolmogorov spectrum in the inertial range) and random motions in the viscous range of scales.

\begin{figure}
\vspace*{2mm} \centering
\includegraphics[width=8cm]{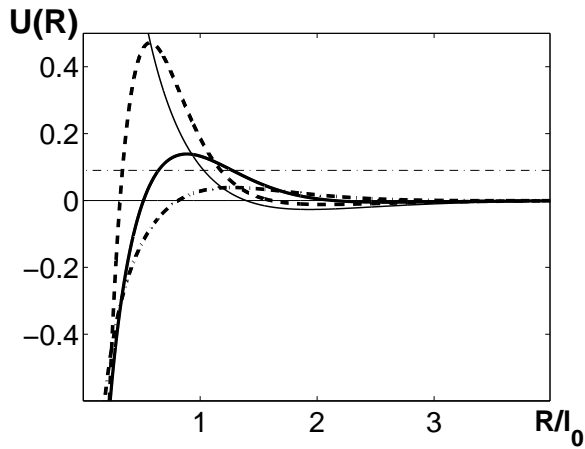}
\caption{\label{Fig9} Potential $U$ versus $R/l_0$ for  $\sigma_{_{T}}=1$ and different values of parameter $a_c$: $\; a_c=1.2$ (dashed-dotted line),  $a_c=2$ (thick solid line) and $a_c=4$ (dashed line). For comparison the case $\sigma_{_{T}}=0$ (the delta-correlated in time velocity field) is shown (thin solid line). Here Re$=2 \times 10^3$ and Pe$=2 \times 10^8$. Horizontal dashed-dotted line $U=0.09$ corresponds to the damping rate for the first mode $\gamma_1 =- 0.09 \, \tau_0^{-1}$.}
\end{figure}

\begin{figure}
\vspace*{2mm} \centering
\includegraphics[width=8cm]{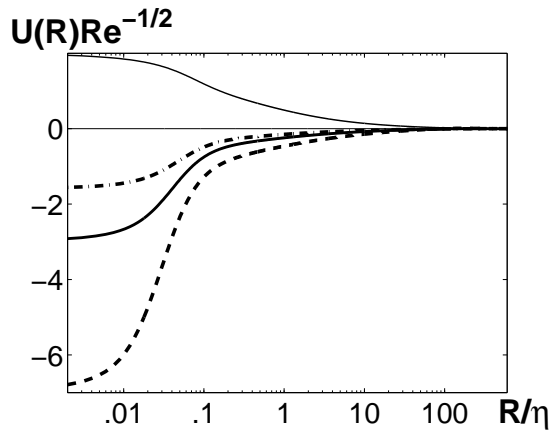}
\caption{\label{Fig10} Potential $U/ {\rm Re}^{1/2}$ versus $R/\eta$ at small scales for  $\sigma_{_{T}}=1$ and different values of parameter $a_c$: $\; a_c=1.2$ (dashed-dotted line),  $a_c=2$ (thick solid line) and $a_c=4$ (dashed line). For comparison the case $\sigma_{_{T}}=0$ (the delta-correlated in time velocity field) is shown (thin solid line). Here $\eta=l_0 / {\rm Re}^{3/4}$ is the Kolmogorov (viscous) scale and $\tau_\eta= \tau_0/ {\rm Re}^{1/2}$ is the Kolmogorov time, Re$=2 \times 10^3$ and Pe$=2 \times 10^8$.}
\end{figure}

\begin{figure}
\vspace*{2mm} \centering
\includegraphics[width=8cm]{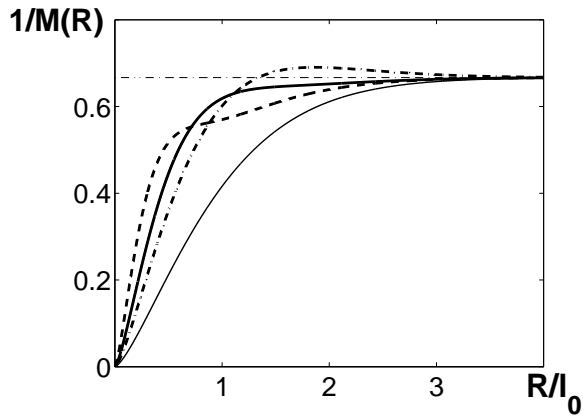}
\caption{\label{Fig11} $1/M$ versus $R/l_0$ for  $\sigma_{_{T}}=1$ and different values of parameter $a_c$: $\; a_c=1.2$ (dashed-dotted line),  $a_c=2$ (thick solid line) and $a_c=4$ (dashed line). For comparison the case $\sigma_{_{T}}=0$ (the delta-correlated in time velocity field) is shown (thin solid line). The horizontal thin dashed-dotted line determined by equation, $M^{-1} =  2/3$, corresponds to the scales at which there is no turbulence. Here Re$=2 \times 10^3$ and Pe$=2 \times 10^8$.}
\end{figure}

Using Eqs.~(\ref{C10})-(\ref{C12}) we determine the potential $U(R)$ and the mass $M(R)$ in Eq.~(\ref{Q4}).  For instance, the potential $U(R)$  for different values of parameter $a_c$ is shown in Fig.~\ref{Fig9}. This potential $U(R)$ for smaller scales is also shown in Fig.~\ref{Fig10}. The function $1/M(R)$  for different values of parameter $a_c$ is shown in Fig.~\ref{Fig11}. Very important features of the potential $U(R)$ which determine the asymptotic behaviour of the solution of Eq.~(\ref{Q4}), are as follows. At small scales, the potential $U(R)$ is negative for $\sigma_{_{T}} \not = 0$, and there is a positive maximum of the potential $U(R)$ at the scales $R \sim 1$. On the other hand, when the degree of compressibility of the turbulent diffusion tensor $\sigma_{_{T}} = 0$ (e.g., for the delta-correlated in time turbulent velocity field), the potential $U(R)$ is positive for all scales and decreases for the larger scales, so that there is no maximum of the potential $U(R)$ for $R > 0$. Analysis performed by Kraichnan in \cite{KR68} for the delta-correlated in time turbulent velocity field, showed that the characteristic damping rate of the particle number density fluctuations is very high $\gamma \sim \tau_\eta^{-1} = \sqrt{\rm Re} \, \tau_0^{-1}$, where $\tau_\eta= \tau_0/\sqrt{\rm Re}$ is the Kolmogorov time (i.e., the turbulent time at the viscous Kolmogorov scale). The latter implies that the level of these fluctuations is very low $\sim (\gamma \,\tau_0)^{-1} \sim {\rm Re}^{- 1/2}$ (see Eq.~(\ref{T32})).

In a real flow with a finite correlation time the degree of compressibility of the turbulent diffusion tensor $\sigma_{_{T}} \not = 0$, the characteristic damping rate of the fluctuations of the particle number density is not high $\gamma \leq \tau_0^{-1}$, and therefore, the level of these fluctuations is not small. In this case the maximum of the potential $U  =  U_{\rm max}$ is attained at the scale $R_{\rm max} \sim 1$, whereby the inverse mass $M^{-1} \sim  1$ and changes with $R$ considerably slower than that in the range $R \ll 1$. The asymptotic solution for the two-point second-order correlation function $\Phi(R)$ in this region has a form of the Gamma distribution $\Phi(R) \propto \exp[-\kappa (R - R_{\rm max})] / R^b$, where $\kappa=\big(U_{\rm max} - 2 \vert \gamma \vert\big)^{1/2}$ and $b \sim 1$. The detailed asymptotic analysis of the solution for the two-point second-order correlation function $\Phi(R)$ in different ranges of scales is performed in Appendix.

\begin{figure}
\vspace*{2mm} \centering
\includegraphics[width=8cm]{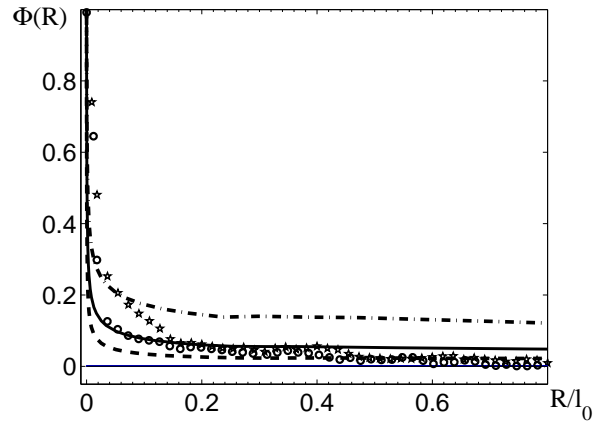}
\caption{\label{Fig12} Normalized two-point second-order correlation function $\Phi(R)$ for  $\sigma_{_{T}}=1$ and different values of parameter $a_c$: $\; a_c=1.2$ (dashed-dotted line),  $a_c=2$ (solid line) and $a_c=4$ (dashed line). Two-point second-order correlation function $\Phi(R)$ determined in our experiments with (unfilled circles) and without (stars) transformation of images into a binary form. The distance $R$ between two points is measured along the $y$ axis at $z=0$ (at the mean boundary of the jet). Here Re$=2 \times 10^3$ and Pe$=2 \times 10^8$.}
\end{figure}

Now we solve Eq.~(\ref{Q4}) numerically in order to determine the two-point second-order correlation function $\Phi_p(R)$. The normalized correlation function for the first mode ($p=1$) for $\sigma_{_{T}}=1$ and different values of parameter $a_c$ is shown in Fig.~\ref{Fig12}. Note that the parameter $a_c$  strongly affects the compressible part of the tensor of turbulent diffusion caused by the finite correlation time of turbulent velocity field. The location of the maximum in the potential $U(R)$ strongly depends on the parameter $a_c$ and is nearly independent of the Reynolds number. The damping rate for the first mode $\gamma_1 =- 0.09 \, \tau_0^{-1}$. The numerical solution for the correlation function $\Phi(R)$ is in agreement with the results of the asymptotic analysis performed in Appendix.

In order to compare the theoretical predictions with the experimental results, the analysis of the experimental data is performed also without transformation of images into a binary form. In particular, the two-point second-order correlation function $\Phi(R)$ determined in our experiments without and with transformation of images into a binary form is shown in Fig.~\ref{Fig12}. The distance between two points $R$ in our experiments is measured along the $y$ axis at $z=0$ (at the mean boundary of the jet). In Fig.~\ref{Fig13} we also show the correlation functions determined in our experiments (without transformation of images into a binary form) for different $z$ inside and outside the jet.

In Figs.~\ref{Fig12} and~\ref{Fig13} we use the Reynolds number, ${\rm Re}$, based on the turbulent velocity velocity $u_0$ and the turbulent correlation length $l_0$, where the characteristic turbulent velocity $u_0 \sim 0.2 V$ and $l_0 \sim D$, so that Re$= 0.2 {\cal R}$. Note that the global Reynolds number ${\cal R}$ based on the mean jet velocity $V$ at the nozzle exit and the diameter $D$ of a jet nozzle, is more often used in experimental studies (e.g., for comparisons of different experimental set-ups), while the Reynolds number, ${\rm Re}$, based on the turbulent velocity velocity $u_0$ and the turbulent correlation length $l_0$, is usually used in theoretical analysis.

\begin{figure}
\vspace*{2mm} \centering
\includegraphics[width=8cm]{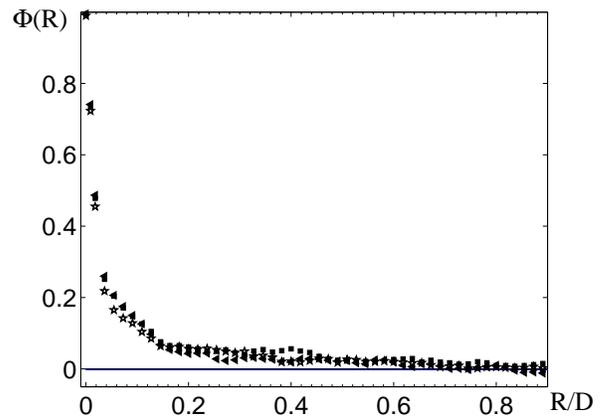}
\caption{\label{Fig13} Two-point second-order correlation function $\Phi(R)$ determined in our experiments without transformation of images in a binary form. The distance $R$ between two points is measured along the $y$ axis: at the mean boundary ($z=0$) of the jet (filled squares); at $z= - 0.55 D$ inside the jet (filled triangles); at $z= 0.55 D$ outside the jet (stars). Here Re$=10^4$ and Pe$=2 \times 10^8$.}
\end{figure}

Figure~\ref{Fig12} demonstrates that a difference between the correlation functions determined in our experiments with and without transformation of images into a binary form is fairly small. Inspection of Fig.~\ref{Fig13} shows that the correlation functions are weakly dependent on $z$. This conclusion is also valid when we transform images into a binary form. Note that when we determine the two-point correlation function using images transformed into a binary form, we take into account the contributions of both, the jet fluid and the ambient fluid. On the other hand, when we determine the parameter $\lambda$ (see Sec.~III), we take into account only the contribution of the jet fluid. The difference between the results obtained using a measured random particle number density field and binarized (using a threshold) random field is not surprising, since a binarization is equivalent to a nonlinear transformation (filtering) of the particle number density field. Note that the two procedures for data analysis of mixing, without and with transformation of images in a binary form, yield an additional information about mixing of particles at the external boundary of a jet. Solid line ($a_c=2$) in Fig. 11 approximates our experimental results after binarization reasonably well.

This study demonstrates that there is a qualitative agreement between the measured and theoretically predicted two-point second-order correlation functions (see Fig.~\ref{Fig12}). This is an indication that the presented theoretical model of the particle number density fluctuations generated by tangling of the gradient of the mean particle number density by the turbulent velocity field, can mimic  mixing at the external boundary of a submerged turbulent jet.

The compressibility of the turbulent diffusion tensor caused by the compressibility of Lagrangian trajectories in a turbulent flow with a finite correlation time is an universal feature of turbulent mixing. The direct consequence of the compressibility is an existence of a positive maximum in the potential $U(R)$ in the scales $R \leq l_0$. This is a reason why the correlation function of the particle number density does not have a power-law behaviour. This property of the correlation function of the particle number density is a generic feature, that seems to be independent of the origin of a random fluid flow. In particular, a complicated random flow which arises at the nonlinear stage of evolution of Rayleigh-Taylor instability has a finite correlation time. This implies that the turbulent diffusion tensor in such flow is compressible. It is plausible to suggest that the latter is a reason that the PDF of the phase function found in the analysis of Rayleigh-Taylor instability in \cite{H06} does not have universal scaling and cannot be approximated by a power law function.

\section{Conclusions}

In this study we investigate experimentally and theoretically mixing at the external boundary of a submerged turbulent jet. An air jet in the experiments is seeded with the incense smoke particles having large Schmidt number and small Stokes number. The spatial distributions of the jet and ambient fluids are determined in these experiments using Particle Image Velocimetry and an Image Processing Technique based on the analysis of the intensity of the light Mie scattering.

Two approaches are used in the data analysis, whereby one approach is based on the measured phase function for the study
of the mixed state of two fluids. This approach is similar to that used for analysis of the Rayleigh-Taylor instability \cite{H06}.
Other approach applied in this study is based on the analysis of the particle number density fluctuations generated by tangling of the gradient of the mean particle number density by the turbulent velocity field. This gradient is formed at the external boundary of a submerged turbulent jet.

The main result of this study is that the phase-function-based formalism and the formalism based on the analysis of the two-point second-order correlation function of the particle number density imply the absence of the universal scaling: the PDF of the phase function of a jet fluid penetrating into an external flow and the two-point second-order correlation function of the particle number density exhibit a non-power-law behaviour.

A power-law behavior of the correlation function of the particle number density can be possible when the characteristic scale of the inhomogeneity of the mean particle number density is larger than the maximum scale of turbulent motions. However, in the vicinity of the jet boundary the characteristic scale of the inhomogeneity of the mean particle number density is much smaller than the maximum scale of turbulent motions. It is plausible to suggest that one of the reasons for the non-universal behavior of the correlation function of the particle number density is the compressibility of Lagrangian trajectories in a turbulent flow with a finite correlation time. For a delta-correlated in time random velocity field the power-law behavior of the correlation function of the particle number density might be possible. However, this suggestion should be carefully verified in different laboratory experiments and numerical simulations.

A mechanism of mixing at the external boundary of a submerged turbulent jet is a kinematic tangling process. The tangling mechanism is universal and independent of the way of generation of turbulence for large Reynolds numbers. Tangling of the gradient of the large-scale velocity shear produces anisotropic velocity fluctuations \cite{L67,WC72} which are responsible for different phenomena: formation of large-scale coherent structures in a turbulent convection \cite{EKRZ02}, excitation of the large-scale inertial waves in a rotating inhomogeneous turbulence \cite{EGKR05}, generation of large-scale vorticity \cite{EKR03} and large-scale magnetic field \cite{RK03} in a sheared turbulence.

Our theoretical study of mixing is based on the analysis of the two-point second-order correlation function of the particle number density. There is a qualitative agreement between theoretical and experimental results obtained in this study. However, the theoretical results obtained in this study cannot be valid in the most general cases since we adopted a number of simplifying assumptions about the turbulence. We considered a homogeneous, isotropic, and incompressible background turbulence.  We assumed also that the generated ''tangling'' fluctuations of particle number density do not affect the background turbulence. The latter assumption implies a one-way coupling between particles and fluid which is reasonable in view of a small mass-loading parameter for particles. We adopted a simplified physical model in order to describe mixing of particles at the turbulent jet interface. In spite of the fact that the simplified model considered in our paper can only mimic mixing in real jets, this model describes robust features of turbulent mixing in our experiments. Clearly, the comprehensive theoretical and numerical studies are required for quantitative description of mixing at the turbulent jet interface.

\begin{acknowledgements}
We thank A.~Krein for his assistance in construction of the
experimental set-up. We also thank I.~Golubev and S.~Rudykh for their
assistance in processing the experimental data.
This research was supported in part by the Israel Science Foundation
governed by the Israeli Academy of Science, the Israeli Universities
Budget Planning Committee (VATAT) and Israeli Atomic Energy
Commission.
\end{acknowledgements}

\appendix

\section{Asymptotic solutions for second-order correlation function}

In this appendix we perform an asymptotic analysis of solution for the two-point second-order correlation function $\Phi(R)$. The correlation function $\Phi(R)$ satisfies the following boundary conditions: $\Phi(R=0) = 1$ and $\Phi(R \to \infty) = 0$. This function has a global maximum at $R=0$ and satisfies the conditions: $\Phi'(R=0) = 0$, $\,\Phi''(R=0) < 0$ and $\Phi(R=0) > |\Phi (R>0)| $.
There are several characteristic regions in the solution for the correlation function $\Phi(R)$: (i) the viscous range $0 \leq R < {\rm Re}^{-3/4}$; (ii) the inertial range ${\rm Re}^{-3/4} \leq R \leq 1$ and (iii) the large scales $R \gg 1$ whereby there is no turbulence. Here $R$ is measured in units of the maximum scale of turbulent motions $l_0$.

The behaviour of the dimensionless two-point second-order correlation function $\Phi(R)$ is described by Eq.~(\ref{T1}). Equation~(\ref{T1}) with $I=0$ has an exact solution in the viscous range $0 \leq R < {\rm Re}^{-3/4}$:
 \begin{eqnarray}
\Phi(X) = S(X) \, X^{-1}  \, (1 + X^2)^{\mu/2} \;,
 \label{T2}
 \end{eqnarray}
where $S(X)$ is a real part of the function $\tilde A_1 P_\zeta^\mu(iX) + \tilde A_2 Q_\zeta^\mu(iX)$. Here $P_\zeta^\mu(i X)$ and $Q_\zeta^\mu(i X)$ are the Legendre functions with imaginary argument, $X^2= \beta_M \, {\rm Pe} \, \sqrt{\rm Re} \, R^2$, and
  \begin{eqnarray*}
\mu &=& {5 \sigma \over 1 + 3 \sigma} \;, \quad \beta_M = {1 + 3 \sigma \over 3(1 +\sigma_{_{T}})} \;,
\\
\zeta &=& - {1 \over 2} \pm {1 \over 2} \biggl[\beta^{2} + 1 - {4|\gamma| \over \beta_M \sqrt{\rm Re}} \biggr]^{1/2}  \;,
\\
\beta &=& {3 - \sigma \over 1 + 3 \sigma} \;,
 \end{eqnarray*}
$\sigma = a_c \, \sigma_{_{T}}$ and $ \gamma < 0$. In order to obtain the solution~(\ref{T2}) we take into account that in the viscous range of scales the inverse mass, the function $\chi(R)$ and the potential $U(R)$ are given by the following formulas:
 \begin{eqnarray*}
M^{-1} &=& {2 \over {\rm Pe}} \, (1 + X^2) \;, \quad \chi(R) = - {2 \beta_U \over 3} \, M(R) \,  {\rm Re}^{1/4} R \;,
\\
U(R) &=& - 2 \beta_U \biggl[1 - \beta_1 \, {X^2 \over 1 + X^2} \biggr] \;,
\\
\beta_U &=& {2 \sigma -1 \over 1 +\sigma_{_{T}}} \;, \quad \beta_1 = {3 (1 + 8 \sigma) \over 1 + 3 \sigma} \; .
 \end{eqnarray*}
Here we use asymptotics of the functions $F(R)$ and $F_c(R)$ determined for $R \ll {\rm Re}^{-3/4}$ (see Eqs.~(\ref{C10})-(\ref{C12}) and ~(\ref{L1})-(\ref{L3})). The solution~(\ref{T2}) has the following asymptotics. In the scales $ 0 \leq R \ll 1 / [\sqrt{\rm Pe}\, {\rm Re}^{1/4}]$ (i.e., for $X \ll 1)$, the solution for the correlation function $\Phi(R)$ reads:
\begin{eqnarray}
\Phi(R) = 1 - {|\gamma| \over 6} \, {\rm Pe} \, R^{2} \;,
\label{T3}
\end{eqnarray}
while in the scales $ 1 / [\sqrt{\rm Pe}\, {\rm Re}^{1/4}]
\ll R < {\rm Re}^{-3/4} $ (i.e., for $X \gg 1)$,
\begin{equation}
\Phi(R) = A_1 + A_2 \, R^{-\beta} \; .
\label{T4}
\end{equation}

In the inertial range ${\rm Re}^{-3/4} \leq R \leq R_{\rm max}$, Eq.~(\ref{T1}) with $I=0$ reads
\begin{eqnarray}
R^2 \Phi'' + (1 - \tilde \beta) \, R \, \Phi' +  {2 \vert \gamma \vert \, \over \tilde \beta_M} \, R^{2/3} \, \Phi = 0 \;,
\label{T10}
\end{eqnarray}
where
\begin{eqnarray*}
\tilde \beta = {3 - \sigma \over 3 + 7 \sigma} \;, \quad \tilde \beta_M = {2 (3 + 7 \sigma) \over 9(1 +\sigma_{_{T}})} \; .
\end{eqnarray*}
Here we use asymptotics of the functions $F(R)$ and $F_c(R)$ determined for ${\rm Re}^{-3/4} \ll R \ll 1$ (see Eqs.~(\ref{C10})-(\ref{C12}) and~(\ref{L1})-(\ref{L2})). Equation~(\ref{T10}) has an exact solution:
 \begin{eqnarray}
\Phi(R) = R^{\tilde \beta/2} \, \big[\tilde A_3 \, J_\nu(z) + \tilde A_4 \, Y_\nu(z)\big] \;,
 \label{T5}
 \end{eqnarray}
where $J_\nu(z)$ and $Y_\nu(z)$ are the Bessel functions of the first and the second kinds, $z = \beta_2 \, R^{1/3}$, $\nu=(3/2) \, |\tilde \beta|$ and $\beta_2 = 3 \big(2 |\gamma| / \tilde \beta_M \big)^{1/2}$. Here we take into account that in the inertial range the inverse mass $M^{-1}(R) = \tilde \beta_M \, R^{4/3}$ and the function $\chi(R) = -(4\beta_U /9) M(R) \, R^{1/3}$.
The solution (\ref{T5}) has the following asymptotics
in the scales ${\rm Re}^{-3/4} \leq R \ll \beta_2^{-1/3}$ (i.e., for $z \ll 1)$:
 \begin{eqnarray}
\Phi(R) = A_3 + A_4 \, R^{-\tilde \beta} \; .
 \label{L5}
 \end{eqnarray}

In the vicinity of the maximum of the potential $U  =  U_{\rm max}$ that is attained at the scale $R_{\rm max} \sim 1$, the inverse mass $M^{-1} \sim  1$ and it changes with $R$ considerably slower than in the range $R \ll 1$. The asymptotic solution for the correlation function $\Phi(R)$ in this region reads:
 \begin{eqnarray}
\Phi(R) = A_5 {\exp\big[-\kappa (R - R_{\rm max})\big] \over R^b} \;,
\label{T8}
 \end{eqnarray}
where $\kappa=\big(U_{\rm max} - 2 \vert \gamma \vert\big)^{1/2}$ and $b \sim 1$.

For large scales, $R \gg 1$, there is no turbulence so that $M^{-1} \sim  2/3$ and $\chi=U=0$.  The asymptotic solution for the correlation function $\Phi(R)$ in this range of scales reads:
 \begin{eqnarray}
\Phi(R) = A_{6} R^{-1} \, \sin \big(R \, \sqrt{3 |\gamma| / 2} + \varphi_0\big) \; .
\label{L6}
 \end{eqnarray}

Matching functions $\Phi(R)$ and $\Phi'(R) $ at the boundaries of the above-mentioned regions yields coefficients $A_{k}$ and the damping rate $\gamma$. In particular, $A_1\sim A_3 \sim A_5 \sim A_6 \approx 1$, and $A_2 \ll 1$, $\, A_4 \ll 1$, where we consider the case Sc$\gg 1$. The results of the above asymptotic analysis are in agreement with the numerical solution for the correlation function $\Phi(R)$ obtained in Sec.~IV.

\end{document}